\documentclass[ 10pt]{article}
\usepackage{amsmath}
\usepackage{amssymb}
\usepackage{amsthm}

\newtheorem{theorem}{Theorem}
\newtheorem{definition}{Definition}

\begin{document}

\title{Undecidable classical properties of observers}
\author{Sven Aerts \\
\textit{Center Leo Apostel for Interdisciplinary Studies (CLEA)} \\
\textit{and Foundations of the Exact Sciences (FUND) } \\
\textit{Department of Mathematics, Vrije Universiteit Brussel} \\
\textit{Pleinlaan 2, 1050 Brussels, Belgium} \\
email: \textsf{saerts@vub.ac.be}}
\maketitle

\abstract{\noindent  A property of a system is called actual, if the observation of the test that pertains to that property, yields an affirmation with certainty.  
We formalize the act of observation by assuming that the outcome correlates with the state of the observed system and is codified as an actual property of the state 
of the observer at the end of the measurement interaction.  For an actual property, the observed outcome has to affirm that property 
with certainty, hence in this case the correlation needs to be perfect.  A property is called classical if either the property or its negation is actual.  
It is shown by a diagonal argument that there exist classical properties of an observer that he cannot observe perfectly.  
Because states are identified with the collection of properties that are actual for that state, it follows that no observer can perfectly observe his own state.  
Implications for the quantum measurement problem are briefly discussed. 
\\ {\bf pacs}: 02.10.-v, 03.65.Ta {\bf keywords}: undecidability, observation, self-reference, quantum measurement 
problem}

\section{Introduction: states and properties}

In the realm of quantum physical experiments, data often consists of nothing
more than clicks of detectors or spots on a photographic plate collected at
certain instances and positions. With a sufficient amount of such clicks,
obtained under specific circumstances, we can establish the properties and
hence determine the state of the system, at least to a precision that
depends in principle only on the amount of clicks we care to gather. How
does this work? The state, a representation of the `mode of being' of a
system, determines the properties of the system. Loosely speaking, a
property of a system is something that can be attributed to that system with
a certain persistency. Birkhoff and von Neumann made the first decisive step
towards the characterization of a system through its properties. They called
quantities which yield only one of two possible results, and which are
testable in a reproducible sense `experimental propositions'. Such
experiments also form the operational basis of the Geneva-Brussels approach,
in which a property is introduced as an equivalence class of tests (also
called questions or experimental projects), and a state is regarded as a set
of actual properties. We will introduce the notion of property and state
through the state-property spaces as can be found in {\cite{Aerts}}, but
this paper is fully self-contained and we will introduce the required
definitions along the way. Consider a physical entity $S$ with its
(non-empty) set of states $\Sigma _{S}$ and (non-empty) set of properties $%
\mathcal{L}_{S}$.\ States will be denoted by small roman script {$p,q,r,$%
\ldots } and properties by small bold script $\mathbf{a}$,$\mathbf{b}$,$%
\mathbf{c}$,... We assume the state fully characterizes the properties of
the entity, some of which may be actual, and some not. What do we mean when
we say an entity has an actual property $\mathbf{a}$? We assume that for
every property {$\mathbf{a}$ }in $\mathcal{L}_{S}$ there exists a test, let
us call it{\ $\hat{\chi}_{\mathbf{a}},$} that {tests property $\mathbf{a}$}.
The result of the test is a simple `yes' or `no'\footnote{%
If one prefers a mathematical definition, one can regard the test as a
mapping $\Sigma _{S}\times \mathcal{L}_{S}\rightarrow \{yes,no\}.$ This is
not customary in the Geneva-Brussels approach, where the test is mainly
regarded as an operational primitive to come to the formal description
provided by the Cartan map and its inverse.}. If the test yields `yes' with%
\emph{\ certainty}, we say property $\mathbf{a}$ is \emph{actual}.

\begin{definition}[actual property]
An entity $S$ in the state{\ $p,$} is said to have \emph{an actual property} 
$\mathbf{a}$ iff the test{\ $\hat{\chi}_{\mathbf{a}}$} that tests property $%
\mathbf{a}$ yields `yes' with certainty.
\end{definition}

Note that this is in accordance with what we intuitively mean when we say
that \textquotedblleft a system has a property\textquotedblright . Two tests 
$\hat{\chi}_{\mathbf{a}}$ and $\hat{\chi}_{\mathbf{b}}$ are called \emph{%
equivalent} when it is the case that for any state $p$ where $\hat{\chi}_{%
\mathbf{a}}$ gives with certainty `yes', we also have that $\hat{\chi}_{%
\mathbf{b}}$ gives with certainty `yes', and vice versa. We will say
equivalent tests, test the same property. Vice versa, a property is
identified with an equivalence class of tests. A property is actual when any
test in the equivalence class yields `yes' with certainty, because all tests
will then yield `yes' with certainty.

A given property $\mathbf{a}$\textbf{\ }$\in \mathcal{L}_{S}$ may be actual
for some of the states{\ $p\in \Sigma _{S}$} of the entity, but not
necessarily for all. To make this notion precise, employing the usual
notation $\mathcal{P}(\Sigma _{S})$ for the set of all subsets of $\Sigma
_{S}$, we postulate the map $\kappa_{S} :\mathcal{L}_{S}\rightarrow \mathcal{%
P}(\Sigma _{S}),$ called the Cartan map, such that $\kappa_{S}(\mathbf{a})$
is the set of states in $\Sigma _{S}$ for which the property $\mathbf{a}$ is
actual$.$ The triple $(\Sigma _{S},\mathcal{L}_{S}, \kappa_{S})$ will be
called a \textit{state property {space}} \cite{Aerts} and fully
characterizes what can be known about the entity with certainty.

\begin{definition}[state property space]
The triple $(\Sigma _{S},\mathcal{L}_{S},$ $\kappa _{S})$, called a \textit{%
state property space}, consists of two sets $\Sigma _{S}$ and $\mathcal{L}%
_{S}$ (where $\Sigma _{S}$ is the set of states of a physical entity $S$,
and $\mathcal{L}_{S}$ its set of properties), and a function $\kappa _{S}:%
\mathcal{L}_{S}\rightarrow \mathcal{P}(\Sigma _{S})$, such that $\mathbf{a}$%
\textbf{\ }$\in \mathcal{L}_{S}$ is actual for the entity in a state{\ $p$}
iff {$p$} $\in $ $\kappa _{S}(\mathbf{a})$.
\end{definition}

When a property is not actual for the entity $S$ in a given state, we will
say this property\textbf{\ }is\emph{\ potential}.

\begin{definition}[potential property]
If, for a given entity $S$ in the state ${p},$ the property $\mathbf{b}\in 
\mathcal{L}_{S}$ is not actual (${p}$\textbf{\ }$\notin $ $\kappa _{S}(%
\mathbf{b})$), then the property $\mathbf{b}$ is called\emph{\ potential}
for the entity $S$ in the state ${p}.$ We write ${p}\in \kappa _{S}^{{C}}(%
\mathbf{b})\equiv \Sigma _{S}-\kappa _{S}(\mathbf{b}).$
\end{definition}

Operationally speaking, there is an obvious inverse of a test that we
introduce in the following definition.

\begin{definition}[inverse test]
If $\hat{\chi}_{\mathbf{a}}$ tests property $\mathbf{a},$ then switching the
roles of `yes' and `no' defines an new test denoted $\hat{\chi}_{\mathbf{a}%
^{\bot }}$, which yields `yes' when $\hat{\chi}_{\mathbf{a}}$ yields `no',
and vice versa. We will call $\hat{\chi}_{\mathbf{a}^{\bot }}$ the \emph{%
inverse test} of $\hat{\chi}_{\mathbf{a}}.$
\end{definition}

The test $\hat{\chi}_{\mathbf{a}^{\bot }}$ is operationally well-defined by
simply switching the outcomes of the test $\hat{\chi}_{\mathbf{a}}.$ The
notation $\hat{\chi}_{\mathbf{a}^{\bot }}$ suggests that it tests
\textquotedblleft the property $\mathbf{a}^{\bot }$\textquotedblright\ , but
in general we cannot associate $\mathbf{a}^{\bot }$ with a well-defined
property as an equivalence class of tests. The problem is that one can
easily give examples of tests that are equivalent to $\hat{\chi}_{\mathbf{a}%
},$ but for which the inverse test is not equivalent to $\hat{\chi}_{\mathbf{%
a}^{\bot }}.$\footnote{%
As an example, take a spin 1 system, for which the spin $S$ can take either
one of three values $\{-1,0,+1\}.$ Call $\mathbf{a}$ the property of
yielding $S=+1$ with certainty when tested, and likewise $\mathbf{b}$ and $%
\mathbf{c}$ the properties of having $S=0$ and $S=-1$ respectively. Because
the actuality of the properties $\mathbf{a}$\ and $\mathbf{b}$\ mutually
exclude each other, we could perhaps propose $\mathbf{b}=\mathbf{a}^{\bot }$%
. By the same token, $\mathbf{c}$\ is a candidate for $\mathbf{b}^{\bot }$.
However, the tests pertaining to $\mathbf{a}$ and $\mathbf{c}$\ are
manifestly not equivalent.} There is however an important class of
properties, the classical properties, for which this problem does not arise.

\begin{definition}[classical property]
A test that has a predetermined answer, is a classical test. A property that
is defined as an equivalence class of classical tests, will be called a 
\emph{classical property}.
\end{definition}

\begin{definition}[inverse classical property]
Given an entity $S$ with a classical property $\mathbf{a}\in \mathcal{L}%
_{S}. $ Then the inverse test $\hat{\chi}_{\mathbf{a}^{\bot }},$ defines a
property denoted $\mathbf{a}^{\bot },$ called the \emph{inverse} of $\mathbf{%
a}$.
\end{definition}

To show that this definition makes sense, we show that for a classical
property, the inverse property $\mathbf{a}^{\bot }$ is well-defined by the
equivalence class of tests that contains $\hat{\chi}_{\mathbf{a}^{\bot }}$.
That is, we have to show that for two arbitrary tests {$\hat{\chi}_{\mathbf{a%
}}$ and $\hat{\chi}_{\mathbf{a}}^{\prime }$} in the equivalence class of a
classical property $\mathbf{a}$, the corresponding inverse tests, {$\hat{\chi%
}_{\mathbf{a}^{\bot }}$ and $\hat{\chi}_{\mathbf{a}^{\bot }}^{\prime },$}
are also equivalent. First note that, because {$\hat{\chi}_{\mathbf{a}}$ }%
and $\hat{\chi}_{\mathbf{a}}^{\prime }$ are classical tests, so are their
inverses, hence all tests give either `yes' or `no' with certainty. Suppose
that, for a given entity in a given state, $\hat{\chi}_{\mathbf{a}^{\bot }}$
gives `yes', then $\hat{\chi}_{\mathbf{a}},$ gives `no'. Since {$\hat{\chi}_{%
\mathbf{a}}$ and $\hat{\chi}_{\mathbf{a}}^{\prime }$} are equivalent by
assumption, {$\hat{\chi}_{\mathbf{a}}^{\prime }$ }would give `no' too. But
then {$\hat{\chi}_{\mathbf{a}^{\bot }}^{\prime }$} gives `yes', so that {$%
\hat{\chi}_{\mathbf{a}^{\bot }}$ and $\hat{\chi}_{\mathbf{a}^{\bot
}}^{\prime }$ }are equivalent, and $\mathbf{a}^{\bot }$ is well-defined.

Obviously, if the property $\mathbf{a}$ is classical, then so is $\mathbf{a}%
^{\bot }$. Equally obvious, we have for a classical property that the
operation of inversion is idempotent: $\mathbf{a}^{\bot \bot }=\mathbf{a}$.
In the light of the preceding remarks, it is natural to postulate that, for
an entity $S$ for which $\mathbf{a}\in \mathcal{L}_{S}$ is a classical
property, we always have that $\mathbf{a}^{\bot }\in \mathcal{L}_{S}$ too.
For an entity $S$ that has a classical property $\mathbf{a}$ , we have that $%
\forall s\in \Sigma _{S}$ either $\mathbf{a,}$ or $\mathbf{a}^{\bot }$ is
actual. Hence a classical property $\mathbf{a}$ partitions the state space $%
\Sigma _{S}$ in just two sets 
\begin{eqnarray}
\kappa _{S}(\mathbf{a})\cup \kappa _{S}(\mathbf{a}^{\bot }) &=&\Sigma _{S}
\label{classical partitioning} \\
\kappa _{S}(\mathbf{a})\cap \kappa _{S}(\mathbf{a}^{\bot }) &=&\varnothing 
\notag
\end{eqnarray}%
We note that in the development of the Geneva-Brussels approach, further
axioms are imposed on $\mathcal{L}_{S}$ so that the properties of a general
physical system form a complete, atomistic and orthomodular lattice \cite%
{Aerts}. The problem of the `inverse property' is then treated by the
introduction of ortho-axioms on $\mathcal{L}_{S}$ \cite{AertsDeses}, and the
role of the inversion is played by the orthocomplementation. For a Boolean
sub-lattice of $\mathcal{L}_{S}$, or for an entity that is classical, (i.e.
an entity for which $\mathcal{L}_{S}$ is a Boolean lattice), the
orthocomplementation reduces to the Boolean NOT. In either case, it is
assumed that, if $\mathbf{a}\in \mathcal{L}_{S},$ then $\mathbf{a}^{\bot
}\in \mathcal{L}_{S}$ too. For our purposes here, we will only require the
existence of the inverse of every \emph{classical} property in $\mathcal{L}%
_{S}$, as established operationally through the definition of the inverse
test.

\section{The formalization of observation}

Up to now, we are in close accordance with the Geneva-Brussels approach 
\cite{Aerts}, which is recognized for being both realistic (entities \emph{are}
in a state that completely describes the status of all properties that
pertain to the entity) and operational (properties are defined as
equivalence classes of experiments). However, the question of \emph{how} to
test a property, is not formalized. According to the Geneva-Brussels
prescription, to see whether property $\mathbf{a}$ is actual, an observer
needs to perform the test $\hat{\chi}_{\mathbf{a}}$. But if states are
indeed realistic descriptions of systems, then in a more detailed account,
the observer has to be regarded as a system having properties in its own
right. We can define the state property space for the observer just as we
did for the entity: $(\Sigma _{M},\mathcal{L}_{M},$ $\kappa _{M})$. The
observing system performs the test and formulates the outcome. This outcome
defines in a natural way an actual property of the state of the observer
after the observation. Indeed, \emph{if} an outcome occurs, the state of the
observer has either the property that $M$ formulates the outcome `yes', or $%
M $ formulates the outcome `no'. We define the property $\mathbf{i}$ as
follows: 
\begin{equation*}
\mathbf{i:}\text{ the outcome indicated by the observing system is `yes'}
\end{equation*}%
If the experiment is not repeated and we re-read the outcome of the
observation, then we need to get the same result. So the outcome of a test
defines a \emph{classical} property of the observer. The particular outcome
of an observation then depends on whether the state of the observer after
the measurement interaction belongs to

\begin{equation}
\kappa _{M}(\mathbf{i})\text{ or to }\kappa _{M}(\mathbf{i}^{\bot })
\label{observer state partitioning}
\end{equation}%
If we call $\Sigma _{M}^{\prime }$ the space of observer states after
interaction, then the indicator property partitions $\Sigma _{M}^{\prime }$ $%
:\kappa _{M}(\mathbf{i})\cap \kappa _{M}(\mathbf{i}^{\bot })=\varnothing $
and $\kappa _{M}(\mathbf{i})\cup \kappa _{M}(\mathbf{i}^{\bot })=\Sigma
_{M}^{\prime }.$ When the observing system is for example a photomultiplier
or a Geiger-Muller counter, we even have $\Sigma _{M}^{\prime }=\Sigma _{M}$
as long as the experiment is running. In an arbitrary small time interval of
the running experiment, the detector has fired, or it hasn't. Hence its
state indicates `yes' or `no' for that interval. This shows that the set of
states the active detector can attain, equals the set of states that
indicate a `yes' or a `no'.

\subsection{The observer interacts to test a property}

So far we have described a system can have properties, which partitions its
set of states, and the observer is the one who indicates the outcome, which
in turn partitions his set of states (\ref{observer state partitioning}).
These are two very basic desiderata of the process of observing a property.
But we have yet to include the most important ingredient: the result of the
observation should pertain to the entity under study. If the observer is
faithfully observing the result of the test corresponding to an actual
property $\mathbf{a}$ of the entity $S$, then \emph{a fortiori} the
observation has to yield `yes' when the property holds. Hence the state of
the observer $m^{\prime }$ after the act of observation has to express `yes':%
\begin{equation}
s\in \kappa _{S}\left( \mathbf{a}\right) \Rightarrow m^{\prime }\in \kappa
_{M}\left( \mathbf{i}\right)  \label{tentative link}
\end{equation}%
If the observer did not interact with the entity, this implication cannot
reasonably be expected to hold. Hereafter, we denote the state of the system 
$S$ under investigation by $s$ and the state of the observer $M$ that is
measuring $S,$ by $m.$ Assume then an observer in a state $m\in \Sigma _{M}\ 
$interacts with an entity in the state $s$ $\in $ $\Sigma _{S}$. There are
many ways conceivable to form a new state from two interacting states, but
we need not go into details. We only assume that the compound state is a
function $\tau $ of the two constituting states:%
\begin{equation}
\tau :\Sigma _{S}\times \Sigma _{M}\rightarrow \Sigma _{S+M}
\label{tau domain and range}
\end{equation}%
From this compound system, there has to be a flow of information to the
state of the observer. After the interaction the state of the observer
(regarded again as a single system) should convey the outcome of the
observation. That is, there has to exist a restriction $\rho $ of the state
of the total system to the set of states of the observer%
\begin{equation}
\rho :\Sigma _{S+M}\rightarrow \Sigma _{M}  \label{restriction}
\end{equation}%
The nature of the restriction $\rho $ is also quite irrelevant for our
purposes. It can be a partial trace, or perhaps a projection onto some
subspace. It is this new state $m^{\prime }$ of the observer that indicates
`yes' or `no', depending on whether it belongs to $\kappa _{M}(\mathbf{i})$
or to $\kappa _{M}(\mathbf{i}^{\bot }).$ We formulate this as $\phi :\Sigma
_{M}\rightarrow \{yes,no\}$ \ 
\begin{eqnarray}
\phi (m^{\prime }) &=&yes\iff m^{\prime }\in \kappa _{M}\left( \mathbf{i}%
\right)  \label{indicator mapping} \\
\phi (m^{\prime }) &=&no\iff m^{\prime }\in \kappa _{M}\left( \mathbf{i}%
^{\bot }\right)  \notag
\end{eqnarray}%
We summarize our model of observation by composing (\ref{indicator mapping}%
), (\ref{restriction}) and (\ref{tau domain and range}) into a single
mapping $o=\phi \circ \rho \circ \tau :$%
\begin{equation}
o:\Sigma _{S}\times \Sigma _{M}\rightarrow \{yes,no\}
\label{outcomes mapping}
\end{equation}%
We assume that $o(s,m)$ is surjective (both `yes' and `no' can be obtained
by an interaction between $S$ and $M$), and that $o(s,m)$ is defined for all%
\footnote{%
We feel justified in this assumption, not because all interactions
necessarily lead to well-defined outcomes, but because outcomes that are not
defined, cannot express a scientific value.} of $\Sigma _{S}\times \Sigma
_{M}$. Besides these two relatively mild regularity conditions, it is
crucial that $o$ is single-valued for the couples $(s,m)$ such that $m$ is a
state that is to measure a property that is \textit{actual} for the system
in the state $s$. Otherwise the implication (\ref{tentative link}) cannot
hold, because actuality means the corresponding test yields `yes' with
certainty. For a potential property the outcome can be either `yes' or `no',
so that in this case $o(s,m)$ could be two-valued in principle. Hence we
propose the following definition of `perfectness':

\begin{definition}[Perfect Observation]
Let $\Lambda \in \mathcal{P}(\mathcal{L}_{S})$ be a non-empty collection of
properties of an entity $S.$ A state $m$ of an observer $M,$ observing an
entity $S$, will be called $\Lambda $-\emph{perfect} iff there exists a
mapping (\ref{outcomes mapping}) \ $o:\Sigma _{S}\times \Sigma
_{M}\rightarrow \{yes,no\},$ such that for each property $\mathbf{a}\in
\Lambda $ and \emph{\ for every} state $s\in \Sigma _{S},$ we have that for
that particular state $m\in \Sigma _{M}$:%
\begin{eqnarray}
s &\in &\kappa _{S}(\mathbf{a})\implies o(s,m)=\text{yes}
\label{perfect observation} \\
s &\in &\kappa _{S}^{C}(\mathbf{a})\implies o(s,m)\text{ }\emph{could}\text{
be yes and}\emph{\ could}\text{ be no}
\end{eqnarray}
\end{definition}

We remark that the particular state $m$ of\ a $\Lambda $-perfect observer%
\emph{\ }$M$ for which the correlation (\ref{tentative link}) holds, will
also be called $\mathbf{\Lambda }$-perfect. This means that an observer $M$ 
\emph{can} be $\mathbf{\Lambda }$-perfect only if there exists at least a
single state $m\in \Sigma _{M}$ that is $\mathbf{\Lambda }$-perfect, and
that he \emph{is} $\mathbf{\Lambda }$-perfect only if he \emph{is} in the
state that realizes (\ref{tentative link}). Moreover, if the state $m$ is $%
\Lambda $-perfect with $\mathbf{a}\in \Lambda $ and we want to concentrate
on this specific property $\mathbf{a}$, then, with slight abuse of notation,
we will write \textquotedblleft $m$ is $\mathbf{a}$-perfect%
\textquotedblright . We argue that this is the most simple case that
deserves to be called perfect observation: when the observer is in a state
that is able to properly observe \emph{at least a single property} of an
entity, \emph{regardless} the particular state the entity is in. Suppose we
made the notion of $\Lambda $-perfectness dependent on the state of $S$,
then the fixed observer, indicating a permanent `yes' (or a `no'), would be
\textquotedblleft perfect\textquotedblright\ in perhaps as much as half the
cases, without even having to bother about making an observation. We argue
that the purpose of observation is to infer the state from the outcomes. The
observer cannot, in general, be expected to know in advance which state he
is measuring, and \textquotedblleft perfectness\textquotedblright\ entails
that he produces the right correlation (\ref{tentative link}) for\emph{\ any 
}state of the entity in its proper state space $\Sigma _{S}$. Of course,
different states $m$ and $m^{\prime }$ can be perfect with respect to
different properties and in this way an observer with a rich state space can
be perfect with respect to a large set of properties, provided he knows how
to attain those states.

Our definition of perfect observation leaves some unnecessary ambiguity with
respect to classical properties. Recall that a property is \emph{classical},
if for any state $s$ of $S,$ we have that either $\mathbf{a}$ or $\mathbf{a}%
^{\bot }$ is actual. Suppose that $\mathbf{a}^{\bot }$ is actual, then $s\in
\kappa _{S}^{C}(\mathbf{a})$ and the definition of perfectness we have now,
only tells us the outcome $\emph{could}$ be yes or$\emph{\ could}$ be no.
But the Geneva-Brussels approach tells us more. First, the result of the
test that corresponds to a classical property is predetermined, so $o(s,m)$
needs to be single-valued: $o(s,m)$ is always no \emph{or}\ is always yes.
Second, the inverse test $\hat{\chi}_{\mathbf{a}^{\bot }}$ was defined by
switching the outcomes of the test $\hat{\chi}_{\mathbf{a}},$ so we know the
result has to be `no' with certainty. Because we formalized the notion of
the observation by means of interaction with an observer, we need to
formalize this `switching' procedure to test the inverse of a classical
property on the level of the state of the observer.

\begin{definition}[Inverse complete]
A mapping $\alpha :\Sigma _{M}\rightarrow \Sigma _{M}$ is called an \emph{%
inversion }(with respect to $o$) iff 
\begin{eqnarray*}
o(s,m) &\neq &o(s,\alpha (m)) \\
o(s,m) &=&o(s,\alpha (\alpha (m)))
\end{eqnarray*}%
A $\Lambda $-\emph{perfect} observer $M$ is called \emph{inverse complete}
iff, for every classical property $\mathbf{a}\in \Lambda $ of $S$ in the
state $s\in \Sigma _{S},$ and for every $m\in \Sigma _{M}$ that tests $%
\mathbf{a}$ perfectly, there exists at least one state $\alpha (m)$ $\in
\Sigma _{M}$.
\end{definition}

Logically speaking, and this is one of the cornerstones in our argument, it
makes no sense to assert that we have observed a classical property to be
actual, when we know that we would have received the same outcome when the
property is not actual. So we argue that the actuality of inverse
properties, requires inverse outcomes: for an observer to be perfect with
respect to a classical property $\mathbf{a}$, there has to exist a state of
the observer that is $\mathbf{a}^{\bot }$-perfect.

\begin{definition}[Classical perfect]
A state $m\in \Sigma _{M}$ of $M$ is $\mathbf{a}$-classical perfect iff $m$
is $\mathbf{a}$-perfect and $\alpha (m)$ exists in $\Sigma _{M}$ which is $%
\mathbf{a}^{\bot }$-perfect. An observer $M$ will be called $\mathbf{a}$%
\emph{-classical perfect} iff he is in a state $m\in \Sigma _{M}$ that is $%
\mathbf{a}$-classical perfect.
\end{definition}

But if an observer is $\mathbf{a}$-perfect and he is able to switch the
roles of `yes' and `no', then he should also be $\mathbf{a}^{\bot }$%
-perfect. We now show this is indeed the case.

\begin{theorem}
Given an entity $S$ with a classical property $\mathbf{a}\in \mathcal{L}_{S}$%
, and an inverse complete\emph{\ }observer $M$ with an inversion $\alpha $.
Then 
\begin{equation*}
m\text{ is }\mathbf{a}\text{-perfect}\iff \alpha (m)\text{ is }\mathbf{a}%
^{\bot }\text{-perfect}
\end{equation*}
\end{theorem}

\textbf{Proof:} We first prove the left to right implication. Because $%
\mathbf{a}\in \mathcal{L}_{S}$ is a classical property, we have that $%
\{\kappa _{S}(\mathbf{a}),$ $\kappa _{S}(\mathbf{a}^{\bot })\}$ is a
partition of $\Sigma _{S}.$ Suppose then first that $s\in \kappa _{S}(%
\mathbf{a})$. If $m$ is $\mathbf{a}$-perfect, then $o(s,m)=$ yes. Hence $%
o(s,\alpha (m))=$ no, indicating $\alpha (m)$ is indeed $\mathbf{a}^{\bot }$%
-perfect when $s\in \kappa _{S}(\mathbf{a})$. Suppose on the other hand that 
$s\in \kappa _{S}(\mathbf{a}^{\bot }).$ By the assumption that $m$ is $%
\mathbf{a}$-perfect, with $\mathbf{a}$ classical, we have that $o(s,m)$ is
single-valued and $o(s,m)$ is always yes \emph{or}\ is always no. By the
definition of the inversion $\alpha $, we then get $o(s,\alpha (m))$ is
always no \emph{or}\ is always yes. Suppose then that $o(s,\alpha (m))$ is
always no. Then $o(s,\alpha (\alpha (m)))$ is always $yes,$ and by the
idempotency of $\alpha ,$ $o(s,m)$ is also always yes. But if $o(s,m)$ is
always yes, then (because $m$ is $\mathbf{a}$-perfect by assumption) this
implies $s\in \kappa _{S}(\mathbf{a}),$ which contradicts the assumption
that $s\in \kappa _{S}(\mathbf{a}^{\bot }).$ Therefore $o(s,\alpha (m))$
cannot be always no, and hence has to be always yes. We then have $s\in
\kappa _{S}(\mathbf{a}^{\bot })\implies o(s,\alpha (m))=yes,$ making $\alpha
(m)$ indeed $\mathbf{a}^{\bot }$\textbf{-}perfect when $s\in \kappa _{S}(%
\mathbf{a}^{\bot }).$ For the reverse implication, call $\mathbf{b}=\mathbf{a%
}^{\bot }$ , and $\alpha (m)=m^{\ast }.$ Because $\mathbf{a}$ is classical,
so is $\mathbf{b}$. Then, by the first part of the proof, we have $m^{\ast }$
is $\mathbf{b}$-perfect implies $\alpha (m^{\ast })$ is $\mathbf{b}^{\bot }$%
-perfect. By idempotency of $\alpha $ we have $o(s,\alpha (m^{\ast
}))=o(s,\alpha (\alpha (m)))=o(s,m)$ and for a classical property $\mathbf{b}%
^{\bot }=\mathbf{a}^{\bot \bot }=\mathbf{a}$. So $m$ is $\mathbf{a}$-perfect.%
$\blacksquare $

As the inversion $\alpha $ is defined on the level of the outcomes generated
by $o(s,m),$ the image $\alpha (m)$ need not necessarily be unique. Still
the last theorem can be reversed, as is shown in the following:

\begin{theorem}
Given an entity $S$ with a classical property $\mathbf{a}\in \mathcal{L}_{S}$
and an observer $M$. If $m\in \Sigma _{M}$ is $\mathbf{a}$-perfect and $%
m^{\ast }\in \Sigma _{M}$ is $\mathbf{a}^{\bot }$-perfect, then $m^{\ast }\ $%
is an inversion of $m$.
\end{theorem}

\textbf{Proof:} Both $m$ and $m^{\ast }$ are in $\Sigma _{M}$, hence $%
m^{\ast }=\beta (m)$ with $\beta :\Sigma _{M}\rightarrow \Sigma _{M}.$ If $m$
is $\mathbf{a}$-perfect and $s\in \kappa _{S}(\mathbf{a}),$ then $o(s,m)$ is
always yes. Hence if $\beta (m)$ is perfectly observing $\mathbf{a}^{\bot }$
then $o(s,\beta (m))$ can never be yes because $\mathbf{a}$ is classical.
Hence we have $o(s,m)\neq o(s,\beta (m)).$ If on the other hand $m$ is $%
\mathbf{a}$-perfect but $s\in \kappa _{S}(\mathbf{a}^{\bot }),$ then $%
o(s,m)=no$. But by assumption $\beta (m)$ is $\mathbf{a}^{\bot }$-perfect
and this is actual, hence $o(s,\beta (m))=yes.$ Because $o(s,m)$ is
two-valued and defined for all of $\Sigma _{S}\times \Sigma _{M}$, the
condition $o(s,m)\neq o(s,\beta (m))$ implies $o(s,m)=o(s,\beta (\beta
(m))). $ So $\beta $ is indeed an inversion.$\blacksquare $

The correlation (\ref{tentative link}) thus becomes much stronger for a
classical property observed by a classically perfect observer. Indeed, if $%
\mathbf{a}$ is classical, we have that either $\mathbf{a}$ or $\mathbf{a}^{%
\mathbf{\bot }}$ is actual. So either $o(s,m)$ or $o(s,\alpha (m))$ has to
be single-valued for any given $s.$ But because $\alpha (m)$ and $m$ are
each others inversion, both $o(s,\alpha (m))$ and $o(s,m)$ have to be
single-valued. So $\mathbf{a}$-classical perfect observation implies 
\begin{eqnarray}
s &\in &\kappa _{S}(\mathbf{a})\iff o(s,m)=yes  \label{classical correlation}
\\
s &\in &\kappa _{S}(\mathbf{a}^{\bot })\iff o(s,m)=no
\label{classical correlation no}
\end{eqnarray}%
Similar equations apply for the state $\alpha (m)$ with yes and no switched.

Arguably, the ultimate purpose of observation is not only to measure a
single property, but to infer the state of an entity by performing tests and
formulating outcomes. Such an observer is able to observe the state of the
entity $S$ and will be called \textquotedblleft $S$-knowledgable
\textquotedblright . In the operational approach to quantum logic \cite%
{Aerts}, the so-called \textquotedblleft state determination
axiom\textquotedblright\ dictates that a state is determined by the set of
properties that are actual in that state. If an observer is to be able to
infer an arbitrary state of $S$, then for each separate actual property $%
\mathbf{a}$ of the entity $S$ that he is asked to observe, he should be able
to acquire a state $m$ that is $\mathbf{a}$-perfect. We formulate this in
our last definition.

\begin{definition}
An observer $M$ will be said to be $S$\emph{-knowledgable}\ iff for an
arbitrary\emph{\ }state $s\in \Sigma _{S}$ of $S$ and for every property $%
\mathbf{a}\in $ $\mathcal{L}_{S}$ that is actual for $s$, there exists a
state $m\in \Sigma _{M}$ that is $\mathbf{a}$-perfect\emph{.}
\end{definition}

Note that this definition is lenient in the sense that it does not require
there should exist a single state of the observer that is perfect for all
properties. However, this advantage automatically disappears when the
observer examines a property of himself, because perfect observation of a
property entails that he gives the right answer, regardless of the state of
the system under investigation!

\subsection{Undecidability in observation}

Can an observer $M$ find out in which state a given system $S$ is? By
definition he can iff for every property $\mathbf{a}$ that is actual in
state $s$, there exists a state $m\in \Sigma _{M}$ that is $\mathbf{a}$%
-perfect.\emph{\ }Obviously, the set of actual properties always includes
the classical properties of $S$, because a property is classical iff the
property or its inverse is actual. Let us concentrate on a specific
classical property $\mathbf{a}$ of $S$. An observer could well be $\mathbf{a}
$-classical perfect, or he might not be. In this way we define a new
candidate property for $M$: the property of \textquotedblleft $\mathbf{a}$%
-classical perfectness\textquotedblright , which we will denote by $\mathbf{p%
}_{\mathbf{a}}$. The test that corresponds to this property -letting the
observer interact with $S$ and verify whether he perfectly observes the
property $\mathbf{a}$- can be reliably performed only by an absolute
observer. But even if such an absolute observer exists, how is the observer
to know for himself whether he is perfect or not with respect to the
observation of a classical property? He cannot rely on the outcome given by
another `god-like' observer, because such an outcome is yet another example
of a classical property that he needs to observe, giving rise to the same
problem. If the observer is to find out, he will have to rely on his own
power of observation. For example, he should at least be sure about what the
outcome he observed, i.e. about the actuality of the classical indicator
property $\mathbf{i}$ . We now show there are fundamental restrictions to
the observation of the classical properties that pertain to himself. 

\begin{theorem}
Let $\mathbf{a}$ be a classical property of $M.$ If $M$ is $\mathbf{a}$%
-classical perfect, then $\mathbf{p}_{\mathbf{a}}$ is classical too.
\end{theorem}

\textbf{Proof:} Because $\mathbf{a}$ is a property of $M,$ and by the
definition of perfect observation, $M$ can only be $\mathbf{a}$-classical
perfect, if $M$ can tell for any state in $\Sigma _{M}$ whether $\mathbf{a}$
is actual or not. Hence all states of $M$ have to be $\mathbf{a}$-classical
perfect. Furthermore, because $\mathbf{a}$ is a classical property, $\mathbf{%
a}$-classical perfect observation entails the outcome has to be
predetermined. A predetermined outcome as a result of an observation of a
classical property, is either always right or always wrong, so that either $%
\mathbf{p}_{\mathbf{a}}$ or $\mathbf{p}_{\mathbf{a}}^{\bot }$ is actual for
all $m\in \Sigma _{M}.\blacksquare $

\begin{theorem}
Let $\mathbf{a}$ be a classical property of $M.$ If $M$ is $\mathbf{a}$%
-classical perfect, then $M$ cannot observe $\mathbf{p}_{\mathbf{a}}$%
-classical perfectly.
\end{theorem}

\textbf{Proof:} We proceed ad absurdum and assume that there exists a
non-empty set of states denoted $\Sigma _{M}^{\mathbf{p}_{\mathbf{a}}},$
such that $m\in \Sigma _{M}^{\mathbf{p}_{\mathbf{a}}}$ observes $\mathbf{p}_{%
\mathbf{a}}$-classical perfect, i.e. $m$ observes\textbf{\ }$\mathbf{p}_{%
\mathbf{a}}$-perfectly, and $\alpha (m)$ observes $\mathbf{p}_{\mathbf{a}%
}^{\bot }$-perfectly. To answer the question whether an arbitrary state $m$
is $\mathbf{p}_{\mathbf{a}}$-classical perfect, $M$ investigates the state $m
$ (either by introspection or by examining an identical system in the same
state). Because the property tested is $\mathbf{p}_{\mathbf{a}}$ -which is
classical by theorem 3- the definition of $\Sigma _{M}^{\mathbf{p}_{\mathbf{a%
}}}$ is obtained by rewriting (\ref{classical correlation}) and (\ref%
{classical correlation no}) for $M,m$ and $\mathbf{p}_{\mathbf{a}}:$

\begin{equation}
\Sigma _{M}^{\mathbf{p}_{\mathbf{a}}}=\{m\in \Sigma _{M}:m\in \kappa _{M}(%
\mathbf{p}_{\mathbf{a}})\iff o(m,m)=yes\}  \label{c correlations m}
\end{equation}

\begin{equation}
\Sigma _{M}^{\mathbf{p}_{\mathbf{a}}}=\{m\in \Sigma _{M}:m\in \kappa _{M}(%
\mathbf{p}_{\mathbf{a}}^{\bot })\iff o(m,m)=no\}  \label{c correlations perp}
\end{equation}%
But classical perfectness with respect to $\mathbf{p}_{\mathbf{a}}$, by
inverse completeness, entails that there exists at least one $\alpha (m)$ in 
$\Sigma _{M}$ with $\alpha (m)\in \kappa _{M}(\mathbf{p}_{\mathbf{a}}^{\bot
}).$ Hence, if $M$ is in the state $\alpha (m),$ application of (\ref{c
correlations perp}) to $\alpha (m)$ gives:%
\begin{equation}
\alpha (m)\in \kappa _{M}(\mathbf{p}_{\mathbf{a}}^{\bot })\iff o(\alpha
(m),\alpha (m))=no  \label{alpha}
\end{equation}

But if equation (\ref{alpha}) holds for $\alpha (m),$ then by (\ref{c
correlations perp}), we have $\alpha (m)\in \Sigma _{M}^{\mathbf{p}_{\mathbf{%
a}}}$. This means $\mathbf{p}_{\mathbf{a}}$ is actual for $\alpha (m)$ and $%
\alpha (m)\in \kappa _{M}(\mathbf{p}_{\mathbf{a}})$. By (\ref{c correlations
m}), the outcome should have been yes.$\blacksquare $

The construction of the proof relies on the necessary existence of $\alpha
(m)$ and is therefore recognized as a diagonal argument. For such an $\alpha
(m)$ we have to assume the outcome is `no', expressing he is not $\mathbf{p}%
_{\mathbf{a}}$-perfect. But then this was a perfect observation and it
should have been `yes'. The structure that we find when an observer attempts
to answer the question of his own non-perfectness, is similar to the
well-known Liar paradox, or the G\"{o}del sentence \textquotedblleft $x:$ x
is not provable\textquotedblright , whose very proof would seem to imply the
truth of the proposition, which states that it is not provable, and so on...
Regarded as a logical proposition, the terminology to indicate this
logically circular decision problem was called `undecidable' by G\"{o}del 
\cite{Godel}, hence the title of this paper. As a consequence of theorem 4,
we now prove that no observer can observe his own state perfectly.

\begin{theorem}
No observer $M$ can be $M$-knowledgable.
\end{theorem}

\textbf{Proof: }An observer is $M$-knowledgable if he can perfectly observe 
\emph{all} actual properties of the state $m$ he is in. Suppose $\mathbf{a}$
is a classical property of $M.$ There is at least one such $\mathbf{a}$,
because we postulated the outcome indicator is a classical property. If $M$
is not $\mathbf{a}$-classical perfect, then he cannot know his own state.
Hence we assume $M$ is $\mathbf{a}$-classical perfect. By theorem 3, $%
\mathbf{p}_{\mathbf{a}}$ is classical too, and he needs to be able to
observe that property classically perfect. We have shown that he cannot,
indicating he cannot observe all his actual properties.\textbf{\ }$%
\blacksquare $

\section{Concluding remarks}

The real problem is, of course, that all observation is self-observation.
The detector doesn't measure an exterior system directly, but rather through
an act of observation in the changes in the state of its own system. Of
course, the argument doesn't deny that real observers \emph{can} make $%
\Lambda $-perfect observations with a high probability of success for a
variety of properties. The result says only that $M$ cannot observe whether
his observation was $\Lambda $-perfect, or not. Also, nothing in our
argument denies the possibility of a second observer observing this first
observer to be perfect (or not!) in his observation. This would preempt the
self-referential loop in the proof. But this second observer faces the same
problem, leading to an infinite regression that cannot solve the original
problem. This reminds one of the way a stronger formal system can be used to
decide whether a given formal system is complete and consistent, but even
this stronger formal system cannot decide its own completeness and
consistency. One could say that the property of perfectness is potential
only. This stance is viable but begs the question how we should observe if
we cannot do it perfectly. Another possible generalization is to allow for a
countable set of outcomes for a test. It is, however, a quite characteristic
feature of undecidability arguments, that the essential result does not
depend on the cardinality of the outcome set, as long as it is countable.
That no observer can observe its own state perfectly, seems in close
accordance with Breuer's results \cite{Breuer1995}, in which Breuer has
shown by a very elegant argument that there exist different states of an
observer that he himself cannot distinguish. To the best of our knowledge,
the first presentation showing the relevance of these issues in relation to
the quantum measurement problem, is the trail-blazing 1977 paper by Dalla
Chiara \cite{DallaChiara1977}. This is not the place to attempt an overview
of the extensive literature on the subject, and we refer to \cite%
{BushLahtiMittelstaedt} and \cite{Svozil} and the references therein. These
results explain why the quantum measurement problem has not been solved. If
the theory describes every fundamental interaction, then it also has to
describe the process of observation. This allows for self-reference because
the theory talks about the way the results are tested, which are produced by
that same theory. Self-reference in turn, allows for the undecidability. An
undecidable proposition cannot be made decidable by additional knowledge
because it is not an epistemological issue. Thus an ontological uncertainty
about at least some of the measurement outcomes becomes unavoidable. In
agreement with \cite{Breuer1995}, we argue this is not necessarily due to
the non-classicality of the properties or theory, but because we regard the
theory as fundamental, i.e. describing \emph{all} processes. Perhaps one
should not describe the process of observation, or describe it in a fashion
entirely different than other processes. This pragmatic stance is taken in
present day quantum physics, and as a result, we have two different,
incompatible evolution laws and no clear rule to tell us what precisely
constitutes an observation and what is a normal interaction, and why they
should be treated differently. It seems we are left with two logical
alternatives\footnote{%
It was kindly pointed out by a referee that a third option would be that a
system does not have actual properties.} for any theory that includes a
description of the observation of the quantities it predicts: either we have
a dichotomic split between the process of observation and other
interactions, or we include both under a single heading and face the
undecidability. In an upcoming article we will argue the second possibility
can serve as an alternative formulation of the basic structure of quantum
probability.

\emph{Acknowledgements}: The author thanks Bart d'Hooghe, Dirk Aerts,
Michiel Seevinck and Freddy De Ceuninck for their valuable comments. This
work was supported by the Fund for Scientific Research (FWO) project
G.0362.03.

\end{document}